\documentclass[12pt]{article}
\usepackage{amsfonts}
\usepackage{pstricks}
\usepackage{pst-node}
\usepackage{epsfig}
\usepackage{amsmath,amssymb,latexsym}

\usepackage{import}

\usepackage{enumerate}

\usepackage{physics}

\usepackage[lmargin=1.0in, rmargin=1.0in, tmargin=1.0in, bmargin=1.0in]{geometry}

\usepackage{cancel}

\usepackage{empheq}

\usepackage{comment}

\usepackage{setspace}
\begin{document}

%--------------------------------------------
% several abbreviations
\def\ba{\begin{eqnarray}}
\def\ea{\end{eqnarray}}
\def\w{\wedge}
%--------------------------------------------

%\setstretch{1.0}

%\begin{titlepage}
\title{\bf Weyl covariance, second clock effect and proper time in theories of symmetric teleparallel gravity}
\author{Caglar Pala$^{1,2,}$\footnote{caglar.pala@gmail.com}, Ozcan Sert$^{2,}$\footnote{osert@pau.edu.tr}, Muzaffer Adak$^{2,}$\footnote{madak@pau.edu.tr} \\
  {\small $^1$ Department of Physics, Faculty of Science, Erciyes University, 38280 Kayseri, Turkey} \\
  {\small $^2$ Department of Physics, Faculty of Science, Pamukkale University, 20017 Denizli, Turkey} \\
 }
 
  \vskip 1cm
\date{12 January 2023}
\maketitle

 \thispagestyle{empty}

\begin{abstract}
 \noindent
Just after Weyl's paper (Weyl in Gravitation und Elektrizität, Sitzungsber. Preuss. Akad., Berlin, 1918) Einstein claimed that a gravity model written in a spacetime geometry with non-metricity suffers from a phenomenon, the so-called second clock effect. We give a new prescription of parallel transport of a vector tangent to a curve which is invariant under both of local general coordinate and Weyl transformations in order to remove that effect. Thus since the length of tangent vector does not change during parallel transport along a closed curve in spacetimes with non-metricity, a second clock effect does not appear in general, not only for the integrable Weyl spacetime. We have specially motivated the problem from the point of view of symmetric teleparallel (or Minkowski-Weyl) geometry. We also conclude that if nature respects Lorentz symmetry and Weyl symmetry, then the simplest geometry in which one can develop consistently alternative gravity models is the symmetric teleparallel geometry; $Q_{\mu\nu}\neq 0, \; T^\mu=0, \; R^\mu{}_\nu=0$. Accordingly we discuss the proper time, the orbit equation of a spinless test body and the Lagrangian for symmetric teleparallel gravity.  \\

%\vskip 1.0cm
%\bigskip


% 11.15.Kc    (Gauge field theory) Classical and semiclassical techniques
% 04.50.Kd    Modified theories of gravity
% 02.40.Yy    Geometric mechanics
\end{abstract}
%\end{titlepage}

%\setcounter{page}{1}

\section{Introduction}

In general relativity (GR) idealised massive spinless test particles moving only under a gravitational field have spacetime histories (orbits) that coincide with timelike geodesic (autoparallel) curves associated with the spacetime metric. On the other hand, in modified theories of gravity formulated in non-Riemannian spacetimes with torsion and/or non-metricity, geodesic and autoparallel curves are not the same and it is not so clear to decide the orbit equation of test body. In Einstein's pseudo-Riemannian description of gravitation the affine parameter, $\tau$, physically is taken as proper time which is measured by a standard clock that is modeled by any timelike curve $x^\mu(\tau)$. Thus in spite of the fact that proper time passed between two events connected by $x^\mu(\tau)$ is path dependent, a standard Einsteinian clock accepts a proper time parametrization independent of its trajectory. Nevertheless, in GR after a parallel transportation of two identical clocks, which are synchronized at the beginning, along different paths, the synchronization disappears. This is known as the first clock effect. On the other hand, in a non-Riemannian geometry containing non-metricity the identification of a clock as a device for measuring proper time requires more care since Hermann Weyl published his influential papers on unification of gravity and electromagnetism \cite{weyl1918},\cite{weyl1919}. Weyl required that the covariant derivative satisfies the semimetricity condition, $\nabla_\sigma g_{\mu\nu}=-2 B_\sigma g_{\mu\nu}$ in his theory where $B_\sigma$ is the Weyl potential vector. Just after his announcement Einstein criticized his theory by saying that rate of a clock must be dependent of its trajectory (history) which is called second clock effect (SCE). Since a SCE has never been observed in laboratory, Weyl's theory fell out of favor over time. In \cite{lobo2018}  observational constraints are set on SCE by investigating recent data on the dilated lifetime of muons accelerated by a magnetic field in CERN.

However, there is a plenty of work containing non-metricity which is performed in the context of modified theories of gravity in literature, e.g. see  \cite{benn1982}-\cite{adak2022} and the references therein. Thus since it is to be worthy to strive for clarifying if there is definitely a SCE or a way for avoiding that in spacetimes with non-metricity, there have been researches \cite{ehlers1972}-\cite{israelquiros2022}. In \cite{ehlers1972} the authors give some axioms and their proofs about a standard clock for measuring proper time in terms of just light rays and freely falling massive particles. Parallel to this work in \cite{perliek1987} the author presents a mathematical characterization of standard clocks in a Weyl manifold which yields an experimental method to test if a given clock is a standard clock or not. In \cite{tucker1996},\cite{tucker1998} the definitions of standard clocks in theories of gravitation formulated in the non-Riemannian spacetime geometries are discussed and it seems possible to define a new standard clock that will be able to measure the affine parameter by considering the invariance of action and the full connection under gauge symmetries. The authors of \cite{avalos2018} find that a Weyl integrable spacetime is the most general Weyl geometry in which a generalized clock defined by Perlick \cite{perliek1987} does not measure a SCE. Additionally, in lemma 2 of reference \cite{koivisto2020} it is stated and proved that the inner product is path-independent iff $R_{(ab)}=0$, and so the authors conclude that a SCE does not arise in symmetric teleparallel spacetimes. Accordingly, for a non-Riemannian geometry not satisfying the condition $R_{(ab)}=0$ a SCE may still be open problem there. In the very recent papers \cite{hobson2020},\cite{hobson2022}  it is argued that a SCE does not occur in Weyl gauge theories of gravity, which are invariant both under local Poincare transformations and local changes of scale. On the other hand, in a different work the authors of \cite{adelhom2020} argue that Perlick's generalized clocks measure a SCE in any non-Riemannian theory of gravity having non-metricity except the form $Q_{\mu\nu}=g_{\mu\nu} d\varphi$ where $\varphi$ is a smooth function. Correspondingly, they state explicitly that generalized clocks in the so-called symmetric teleparallel theories will in general measure a SCE. Similarly, in the recent paper \cite{quiros2021} it is argued that the symmetric teleparallel theories of gravity do not represent phenomenologically viable descriptions of nature due to SCE. Then in a subsequent article \cite{israelquiros2022} the same researcher argues that when only Weyl integrable spacetimes, $Q_{\mu\nu}=g_{\mu\nu} d\varphi$, are considered, the Weyl gauge theories and the symmetric teleparallel gravity theories are free of a SCE. In this work we discuss SCE especially for theories of symmetric teleparallel gravity and conclude that a non-Riemannian spacetime geometry with non-metricity does not need to suffer from second clock effect in general, not only for Weyl integrable spacetimes, by defining a new parallel transport rule of a tangent vector which is invariant under a local general coordinate transformation and a Weyl (conformal or scale) transformation. Our result is not valid only for a symmetric teleparallel spacetime geometry but also for any non-Riemannian geometry containing both non-metricity and torsion. In addition, we arrive at a second conclusion that as long as nature respects Weyl symmetry apart from the Lorentz symmetry, the simplest spacetime geometry in which alternative gravity models can be written is the symmetric teleparallel spacetime defined by $Q_{\mu\nu} \neq 0$, $T^\mu =0$, $R^\mu{}_\nu=0$. In our third conclusion we saw that meta-geodesic curve may be used rather than autoparallel curve in order to represent the orbit of a spinless test body moving under influence of only a gravitating source.  At the end we write down a Lagrangian 4-form invariant under the Lorentz, the Weyl and $U(1)$ transformations which can be a unification theory of gravitational and electromagnetic interactions. Our formalism is developed in terms the language of exterior algebra \cite{thirring1997}, \cite{frankel2012}.

\section{The mathematical preliminaries}

In spite of the fact that we discuss the subject in four dimensions in this work, all the content in this section is valid in any dimensions. The spacetime, in general, is denoted by the triple $\{ M,g,\omega\}$ where $M$ is the $4$-dimensional orientable and differentiable manifold, $g$ is the (0,2)-type symmetric and non-degenerate metric tensor, $\omega$ is the connection 1-form representing the parallel transport of the tensors (and also spinors). We set up a coordinate system $\{x^\mu\}, \ \mu =\hat{0},\hat{1},\hat{2},\hat{3}$, to represent points (events) in $M$. With the abbreviation $\partial_\mu := \{\frac{\partial}{\partial x^\mu}\}$, the set $\{\partial_\mu\}$ denotes a frame. The union of frames constructed on all points of $M$ constitutes the coordinate (holonomic) frame bundle, $CF(M)$, over $M$. We notate the dual of $CF(M)$ by $CF^*(M)$, the so-called coordinate (holonomic) co-frame bundle over $M$. Duality relation is given by $dx^\mu(\partial_\nu)=\delta^\mu_\nu$ where the set $\{dx^\mu\}$ forms a co-frame and $\delta^\mu_\nu$ is the Kronecker symbol. In the language of exterior algebra, the elements of $\{dx^\mu\}$ are called the coordinate (holonomic) 1-forms. We can write the metric tensor in terms of the coordinate co-frame $g=g_{\mu\nu} dx^\mu \otimes dx^\nu$ where $\otimes$ denotes symmetric tensor product, $dx^\mu \otimes dx^\nu =  dx^\nu \otimes dx^\mu$. Thus $g_{\mu\nu}$ is the symmetric coordinate components of metric which is determined by the scalar product of the coordinate co-frame, $g(\partial_\mu , \partial_\nu)=g_{\mu\nu}$. Finally we define connection (or covariant derivative) on any $(p,q)$-type tensor-valued exterior form $\mathfrak{T}^{\mu_1 \mu_2 \cdots \mu_p}_{\; \; \; \; \nu_1 \nu_2 \cdots \nu_q }$ below
 \begin{eqnarray}\label{gl-covariant-derivative}
   D \mathfrak{T}^{\mu_1 \mu_2 \cdots \mu_p}_{\; \; \; \; \nu_1 \nu_2 \cdots \nu_q } = d \mathfrak{T}^{\mu_1 \mu_2 \cdots \mu_p}_{\; \; \; \; \nu_1 \nu_2 \cdots \nu_q }
    + \omega^{\mu_1}{}_\sigma \wedge \mathfrak{T}^{\sigma \mu_2 \cdots \mu_p}_{\; \; \; \; \nu_1 \nu_2 \cdots \nu_q } + \cdots + \omega^{\mu_p}{}_\sigma \wedge \mathfrak{T}^{\mu_1 \mu_2 \cdots \sigma}_{\; \; \; \; \nu_1 \nu_2 \cdots \nu_q } \nonumber \\
     - \omega^\sigma{}_{\nu_1} \wedge \mathfrak{T}^{\mu_1 \mu_2 \cdots \mu_p}_{\; \; \; \; \sigma \nu_2 \cdots \nu_q } - \cdots
    - \omega^\sigma{}_{\nu_q} \wedge \mathfrak{T}^{\mu_1 \mu_2 \cdots \mu_p}_{\; \; \; \; \nu_1 \nu_2 \cdots \sigma} 
 \end{eqnarray}
where $\wedge$ is the exterior product, $d$ is the exterior derivative satisfying the Poincare lemma $d^2=0$, $\omega^\mu{}_\nu$ is the full connection 1-form of $CF^*(M)$ and $D$ is the covariant exterior derivative with respect to $\omega^\mu{}_\nu$. Thus we can write down the Cartan structure equations; non-metricity 1-form, torsion 2-form and full curvature 2-form, respectively,
 \begin{subequations}
  \begin{align}
    Q_{\mu \nu} &:= -\frac{1}{2} D g_{\mu\nu} = \frac{1}{2} (-d g_{\mu\nu} + \omega^\sigma{}_\mu g_{\sigma \nu} +\omega^\sigma{}_\nu g_{\mu \sigma}) \ ,  \label{nonmet-coor}\\
    T^\mu  &:= D dx^\mu = {\omega^\mu}_\nu \wedge dx^\nu \ , \label{tors-coor}\\
    {R^\mu}_\nu &:= D{\omega^\mu}_\nu :=  d {\omega^\mu}_\nu + {\omega^\mu}_\sigma \wedge {\omega^\sigma}_\nu \ , \label{curva-coor}
   \end{align}
 \end{subequations}
where we use the Poincare lemma on the right had side of torsion. They satisfy the Bianchi identities
 \begin{subequations}\label{eq:biancis}
  \begin{align}
   D Q_{\mu\nu} &= \frac{1}{2} ( R_{\mu\nu} +R_{\nu\mu}) \ , \label{bianc:0} \\
       D T^\mu    &= {R^\mu}_\nu \wedge dx^\nu \ ,  \label{bianc:1} \\
       D {R^\mu}_\nu &= 0 \ . \label{bianc:2}
   \end{align}
 \end{subequations}
The spacetime geometry is classified with respect to vanishing of $Q_{\mu\nu}$, $T^\mu$ and $R^\mu{}_\nu$ as shown in Table \ref{tab:spacetimes}.

 \begin{table}[h!]  
   \centering
  \caption{Classification of spacetime. In literature, sometimes firstly $Q_{\mu\nu}$ is decomposed as $Q_{\mu\nu}=\overline{Q}_{\mu\nu}+\frac{1}{4}g_{\mu\nu}Q$ where $g^{\mu\nu}Q_{\mu\nu}= Q$ and $g^{\mu\nu}\overline{Q}_{\mu\nu}=0$, then the case of $\overline{Q}_{\mu\nu}=0$ and $Q\neq 0$ is called Weyl geometry. But, here by ``Weyl geometry'' we mean $Q_{\mu\nu} \neq 0$ in general.\\}
 \begin{tabular}{|r|c|c|c|l|}
 \hline
  & $Q_{\mu\nu}$ & $T^\mu$ & $R^\mu{}_\nu$ & $Geometry$  \\
 \hline \hline
 1 & $0$ & $0$ & $0$ & Minkowski  \\
  \hline
 2 &  $0$ & $ 0$ & $\neq 0$ & Riemann  \\
  \hline
 3 & $0$ & $\neq 0$ & $ 0$ & Weitzenböck teleparallel  \\
  \hline
 4 & $0$ & $\neq 0$ & $\neq 0$ & Riemann-Cartan  \\
  \hline
 5 & $\neq 0$ & $0$ & $0$ & Symmetric teleparallel  \\
  \hline
 6 & $\neq 0$ & $0$ & $\neq 0$ & Riemann-Weyl \\
  \hline
 7 &  $\neq 0$ & $\neq 0$ & $ 0$ & General teleparallel  \\
  \hline
 8 & $\neq 0$ & $\neq 0$ & $\neq 0$ & Most general \\
  \hline
 \end{tabular}
 \label{tab:spacetimes}
  \end{table}

In a coordinate frame the full connection 1-form can be decomposed uniquely as the Riemannian piece plus non-Riemannian piece \cite{benn1982},\cite{tucker1995},\cite{hehl1995}
 \begin{equation}
  {\omega^\mu}_\nu = \widetilde{\omega}^\mu{}_\nu + {\mathrm{L}^\mu}_\nu \label{connec-decomp-coor}
 \end{equation}
where the Levi-Civita (Christoffel) connection 1-form $\widetilde{\omega}^\mu{}_\nu$ is expressed in terms of metric
   \begin{equation}
\widetilde{\omega}^\mu{}_\nu =  \frac{1}{2}g^{\mu \sigma}(\partial_\gamma g_{\sigma \nu} + \partial_\nu g_{\sigma \gamma} - \partial_\sigma g_{\nu \gamma})dx^\gamma
  \end{equation}
and the disformation tensor valued 1-form ${\mathrm{L}^\mu}_\nu$ contains tensors of non-metricity 1-form and torsion 2-form
 \begin{equation}
    \mathrm{L}_{\mu \nu} =   Q_{\mu \nu} + (\iota_\nu Q_{\sigma \mu} - \iota_\mu Q_{\sigma \nu})dx^\sigma   + \frac{1}{2} \left[ \iota_\mu T_\nu - \iota_\nu T_\mu - (\iota_\mu \iota_\nu T_\sigma) dx^\sigma \right]
 \end{equation}
where $\iota_\mu$ is the interior product of the exterior algebra, $\iota_\nu dx^\mu=\delta^\mu_\nu$. In the literature sometimes $T^\mu$ is given by the contortion 1-form $K_{\mu\nu}=-K_{\nu \mu}$ via the expression $K^\mu{}_\nu \wedge dx^\nu = T^\mu$. Under the decomposition (\ref{connec-decomp-coor}) we can split the full curvature 2-form in the equation (\ref{curva-coor}) as the Riemannian curvature 2-form, $\widetilde{R}^\mu{}_\nu= d\widetilde{\omega}^\mu{}_\nu + \widetilde{\omega}^\mu{}_\sigma \wedge \widetilde{\omega}^\sigma{}_\nu$, plus the non-Riemannian pieces
 \begin{equation}
     {R}^\mu{}_\nu = \widetilde{R}^\mu{}_\nu +  \widetilde{D} \mathrm{L}^\mu{}_\nu +\mathrm{L}^\mu{}_\sigma \wedge \mathrm{L}^\sigma{}_\nu
 \end{equation}
where $\widetilde{D}$ denotes the covariant exterior derivative with respect to the Levi-Civita connection 1-form $\widetilde{\omega}^\mu{}_\nu$.

By using vielbein (tetrad) $h^\mu{}_a$ we can pass from a coordinate frame $\{\partial_\mu\}$ to an orthonormal frame $\{X_a\}$ by relation $ X_a = h^\mu{}_a  \partial_\mu$ where $a=0,1,2,3$. Of course, we have the inverse, $\partial_\mu = h^a{}_\mu X_a$ such that $h^b{}_\mu h^\mu{}_a =\delta^b_a$ and $h^\mu{}_a h^a{}_\nu=\delta^\mu_\nu$. Through this work Latin indices are the orthonormal (Lorentz or sometimes anholonomic) indices taking values $0,1,2,3$ and Greek ones are the coordinate (holonomic) values taking values $\hat{0},\hat{1},\hat{2},\hat{3}$. Here orthonormality becomes explicit when we write the metric tensor in the orthonormal (Lorentz) co-frame $\{e^a\}$ as $g=\eta_{ab}e^a \otimes e^b$ where $\eta_{ab}$ is the Minkowski metric with the signature $\eta_{ab} = \text{diag}(-1,+1,+1,+1)$. Then metric components are related through vielbein, $\eta_{ab}=h^\mu{}_a h^\nu{}_b g_{\mu\nu}$ or $g_{\mu\nu}=h^a{}_\mu h^b{}_\nu \eta_{ab}$.  Union of orthonormal frames constructed over all points of $M$ constitutes the orthonormal frame bundle, $OF(M)$, and its dual is the orthonormal co-frame bundle, $OF^*(M)$, over $M$. Duality relation is $e^a(X_b)=\delta^a_b$. Again we can use vielbein for transitions between $\{dx^\mu\}$ and $\{e^a\}$ by relations $e^a=h^a{}_\mu dx^\mu$ or $dx^\mu = h^\mu{}_a e^a$. Correspondingly, while a co-frame transforms from $\{dx^\mu\}$ to $\{e^a\}$ in the form of $e^a=h^a{}_\mu dx^\mu$, in order $D$ to transform covariantly, i.e. $De^a=h^a{}_\mu Ddx^\mu$, the full connection 1-form must transform as follows
 \begin{subequations}
   \begin{align}
        \omega^a{}_b &= h^a{}_\mu \omega^\mu{}_\nu h^\nu{}_b + h^a{}_\mu dh^\mu{}_b , \\
   \omega^\mu{}_\nu &= h^\mu{}_a \omega^a{}_b h^b{}_\nu + h^\mu{}_a dh^a{}_\nu \ .
   \end{align}
 \end{subequations}
Thus in the orthonormal frame we can rewrite non-metricity 1-form, torsion 2-form and the full curvature 2-form, respectively,
   \begin{subequations}
 \begin{align}
  Q_{ab} &:= -\frac{1}{2}D\eta_{ab} = \frac{1}{2} (\omega_{ab} + \omega_{ba}) = h^\mu{}_a h^\nu{}_bQ_{\mu\nu}\ , \label{nonmet-orth}\\
    T^a &:= De^a= d e^a + {\omega^a}_b \wedge e^b = h^a{}_\mu T^\mu \ ,  \label{tors-orth}\\
    {R^a}_b &:= D\omega^a{}_b := d {\omega^a}_b + {\omega^a}_c \wedge {\omega^c}_b = h^a{}_\mu h^\nu{}_b R^\mu{}_\nu \ . \label{curva-orth}
   \end{align}
 \end{subequations}
The Bianchi identities given by (\ref{eq:biancis}) may be rewritten readily in the orthonormal frame if wanted. Now we can again decompose the full connection 1-form, $\omega_{ab}$, as the Riemannian part plus others.
 \begin{equation}
     \omega_{ab}= \widetilde{\omega}_{ab} + \mathrm{L}_{ab}
 \end{equation}
where the Levi-Civita connection 1-form, $\widetilde{\omega}_{ab}$, is obtained from orthonormal co-frame,
 \begin{equation}\label{Levi-Civita:on}
   \widetilde{\omega}_{ab} = \frac{1}{2} \left[ -\iota_a de_b + \iota_b de_a + (\iota_a \iota_b de_c) e^c \right],
 \end{equation}
and the disformation 1-form from non-metricity and torsion
 \begin{equation}
     \mathrm{L}_{ab}=Q_{ab} + ( \imath_b Q_{ac} - \imath_a Q_{bc} ) e^c + \frac{1}{2} \left[ \iota_a T_b - \iota_b T_a - (\iota_a \iota_b T_c) e^c \right].
 \end{equation}

\section{Local general coordinate transformation, Weyl transformation and second clock effect} \label{sec:gen-coor-trans}

In this section firstly we consider the general coordinate transformation (GCT) on $M$ by forgetting the origin
 \begin{equation} \label{gen-lin-coord-trans}
  x^\mu \to  x^{\mu'} = \Gamma^{\mu'}{}_\mu x^\mu + \xi^{\mu'}
 \end{equation}
where $\Gamma^{\mu'}{}_\mu $ represents the rotational part and $\xi^{\mu'}$ translational part. Because of the translational piece, the origins of two coordinate systems, $\{x^\mu\}$ and $\{x^{\mu'}\}$, do not have to coincide. GCT can be written in matrix notation. 
 \begin{equation}
     \begin{pmatrix}
      x^{\hat{0}'} \\
      x^{\hat{1}'} \\
      x^{\hat{2}'} \\
      x^{\hat{3}'} \\
      1
     \end{pmatrix} =
     \begin{pmatrix}
      \Gamma^{\hat{0}'}{}_{\hat{0}} & \Gamma^{\hat{0}'}{}_{\hat{1}} & \Gamma^{\hat{0}'}{}_{\hat{2}} & \Gamma^{\hat{0}'}{}_{\hat{3}} &
      \xi^{\hat{0}'}\\
      \Gamma^{\hat{1}'}{}_{\hat{0}} & \Gamma^{\hat{1}'}{}_{\hat{1}} & \Gamma^{\hat{1}'}{}_{\hat{2}} & \Gamma^{\hat{1}'}{}_{\hat{3}} &
      \xi^{\hat{1}'}\\
      \Gamma^{\hat{2}'}{}_{\hat{0}} & \Gamma^{\hat{2}'}{}_{\hat{1}} & \Gamma^{\hat{2}'}{}_{\hat{2}} & \Gamma^{\hat{2}'}{}_{\hat{3}} &
      \xi^{\hat{2}'}\\
      \Gamma^{\hat{3}'}{}_{\hat{0}} & \Gamma^{\hat{3}'}{}_{\hat{1}} & \Gamma^{\hat{3}'}{}_{\hat{2}} & \Gamma^{\hat{3}'}{}_{\hat{3}} &
      \xi^{\hat{3}'}\\
      0 & 0 & 0 & 0 & 1
     \end{pmatrix}
     \begin{pmatrix}
      x^{\hat{0}} \\
      x^{\hat{1}} \\
      x^{\hat{2}} \\
      x^{\hat{3}} \\
      1
     \end{pmatrix}
 \end{equation}
Here $5\times 5$ transformation matrices form the affine group. In general, GCT may be dependent of coordinates. In other words, we consider a local GCT,  $\Gamma^{\mu'}{}_\mu =\Gamma^{\mu'}{}_\mu (x)$ and $\xi^{\mu'}=\xi^{\mu'}(x)$. Thus, by taking the exterior derivative of a local GCT we obtain the transformation rule in $CF^*(M)$ over $M$
  \begin{equation}
   dx^{\mu'} = \Lambda^{\mu'}{}_\mu(x) dx^\mu \label{eq:gct-ctm}
 \end{equation}
where $\Lambda^{\mu'}{}_\mu(x) := [\partial_\mu \Gamma^{\mu'}{}_\nu (x)]x^\nu + \Gamma^{\mu'}{}_\mu(x) + \partial_\mu\xi^{\mu'}(x)$. These transformation elements can be represented by $4\times 4$ matrices, $\left[ \Lambda^{\mu'}{}_\mu(x)\right]$, and also form the general linear group, $\left[ \Lambda^{\mu'}{}_\mu(x)\right] \in Gl(4,\mathbb{R})$. If the full connection of $CF^*(M)$ transform in the following way
 \begin{equation} \label{eq:gl-trans-connec-coord}
   \omega^{\mu'}{}_{\nu'} = \Lambda^{\mu'}{}_\mu \omega^\mu{}_\nu \Lambda^\nu{}_{\nu'} + \Lambda^{\mu'}{}_\mu d\Lambda^\mu{}_{\nu'}.
 \end{equation}
then, the covariant exterior derivative transform covariantly. Thus the non-metricity 1-form, the torsion 2-form and the full curvature 2-form of $CF^*(M)$ transform covariantly
  \begin{subequations}
  \begin{align}
    Q_{\mu' \nu'} &= \Lambda^\mu{}_{\mu'} \Lambda^\nu{}_{\nu'} Q_{\mu\nu}  \ , \\
    T^{\mu'} &= \Lambda^{\mu'}{}_\mu T^\mu \ , \\
    R^{\mu'}{}_{\nu'} &= \Lambda^{\mu'}{}_\mu R^\mu{}_\nu L^\nu{}_{\nu'} \ . 
  \end{align}
 \end{subequations}

By usage of vielbein we can obtain the transformation rule in $OF^*(M)$ over $M$ generated by a local GCT on $M$
 \begin{equation}
  dx^{\mu'} = h^{\mu'}{}_{a'} e^{a'} \quad \mbox{and} \quad dx^{\mu} = h^{\mu}{}_{a} e^{a}
 \end{equation}
where $ h^{\mu'}{}_{a'} =h^{\mu'}{}_{a'} (x'(x))$ and $h^{\mu}{}_{a}=h^{\mu}{}_{a}(x)$. After substituting these into the equation (\ref{eq:gct-ctm})  and then by using the relation $h^{b'}{}_{\mu'}h^{\mu'}{}_{a'} = \delta^{b'}_{a'}$ we obtain
  \begin{equation}\label{oncoframe-transf}
   e^{a'} = L^{a'}{}_a e^a
  \end{equation}
where $L^{a'}{}_a (x) := h^{a'}{}_{\mu'} h^{\mu}{}_a \Lambda^{\mu'}{}_\mu$. In order to decide which group is formed by $L^{a'}{}_a (x)$ we look at the transformation of metric components 
 \begin{equation}
   g= \eta_{ab}e^a \otimes e^b = \eta_{a'b'} e^{a'} \otimes e^{b'}
  \end{equation}
where $\eta_{ab}=\eta_{a'b'}= \mbox{diag}(-1,+1, +1, +1)$ is the Minkowski metric. Together with $e^{a'} = L^{a'}{}_a e^a$ that yields $\eta_{ab}=L^a{}_{a'} \eta_{ab} L^b{}_{b'}$. In the matrix notation it is read as $\left[\eta_{ab}\right] = \left[ L^a{}_{a'} \right]^T \left[\eta_{ab}\right] \left[ L^b{}_{b'}\right]$
where $^T$ denotes the transpose matrix. This shows that the transformation elements, $L^{a'}{}_a$, obtained from a local GCT generate the Lorentz group $SO(1,3)$ in $OF^*(M)$ \cite{adak2022ijgmmp},\cite{thirring1997},\cite{frankel2012}. That is the reason for $\{e^a\}$ to be called the Lorentzian co-frame. Accordingly, for covariant exterior derivative to transform covariantly the full connection of $OF^*(M)$ must transforms as follows 
 \begin{equation}\label{onconnection-transf}
   \omega^{a'}{}_{b'} = L^{a'}{}_a \omega^a{}_b L^b{}_{b'} + L^{a'}{}_a dL^a{}_{b'}.
 \end{equation}
As a result, under a local GCT the non-metricity 1-form, the torsion 2-form and the full curvature 2-form on $OF^*(M)$ transform covariantly
 \begin{subequations}
  \begin{align}
    Q_{a' b'} &= L^a{}_{a'} L^b{}_{b'} Q_{ab}  \ , \label{onQ-transf}\\
    T^{a'} &= L^{a'}{}_a T^a \ , \label{onT-transf}\\
    R^{a'}{}_{b'} &= L^{a'}{}_a R^a{}_b L^b{}_{b'} \ . \label{onR-transf}
  \end{align}
 \end{subequations}
As a complementary remark we give the transformation of veilbein under a local GCT
 \begin{equation}
     h^{a'}{}_{\mu'}= L^{a'}{}_a h^a{}_\mu \Lambda^\mu{}_{\mu'} .
 \end{equation}

Secondly we summarize the Weyl (scale or conformal) transformation independent of local GCT
 \begin{equation}
     g \to \bar{g}=e^{2\psi(x)}g
 \end{equation}
where $\psi(x)$ is a smooth function.  That corresponds to transformations in the coordinate co-frame bundle and the orthonormal co-frame bundle, respectively,
 \begin{subequations}
  \begin{align}
     dx^{\bar{\mu}} &= dx^\mu  \qquad \, \text{and} \quad  g_{\bar{\mu}\bar{\nu}}=e^{2\psi(x)} g_{\mu\nu}, \label{eq:weyl-trans-coor}\\
     e^{\bar{a}} &= e^{\psi(x)} e^a \quad  \text{and}  \quad \eta_{\bar{a}\bar{b}} = \eta_{ab}.
     \end{align}
 \end{subequations}
Accordingly, the Weyl transformation of vielbein is obtained as 
 \begin{equation}
     h^{\bar{\mu}}{}_{\bar{a}}=e^{-\psi(x)} h^\mu{}_a \quad \text{and} \quad h^{\bar{a}}{}_{\bar{\mu}}=e^{\psi(x)} h^a{}_\mu .
 \end{equation}
Thus the conformal (Weyl) weights are $w(dx^\mu)=0$, $w(g_{\mu\nu})=2$, $w(e^a)=1$, $w(\eta_{ab})=0$, $w( h^\mu{}_a)=-1$, $w(h^a{}_\mu)=1$. As this transformation rescales the lengths of vectors, it leaves the angles between two vectors the same. The transformation element $e^{\psi(x)}$ forms the Weyl group, $\mathcal{W}$. We adhere a notation that a prime and a bar denote a local GCT and a Weyl transformation, respectively. Besides, independently from metric we give $\mathcal{W}$-transformation rule of the full connection of $CF^*(M)$ and $OF^*(M)$, respectively,
 \begin{equation}
    \omega^{\bar{\mu}}{}_{\bar{\nu}}= \omega^\mu{}_\nu \quad \text{and} \quad      \omega^{\bar{a}}{}_{\bar{b}}= \omega^a{}_b - \delta^a_b d\psi  .
 \end{equation}
Thus the non-metricity, the torsion and the full curvature transform in the coordinate co-frame bundle
 \begin{subequations}
  \begin{align}
      Q_{\bar{\mu}\bar{\nu}} &= e^{2\psi(x)} (Q_{\mu\nu} - g_{\mu\nu}d\psi(x)),\\
      T^{\bar{\mu}} &= T^\mu, \\
      R^{\bar{\mu}}{}_{\bar{\nu}} &= R^\mu{}_\nu,
  \end{align}
 \end{subequations}
and in the orthonormal co-frame bundle
  \begin{subequations}
  \begin{align}
      Q_{\bar{a}\bar{b}} &= Q_{ab} - \eta_{ab}d\psi(x),\\
      T^{\bar{a}} &= e^{\psi(x)}T^a, \\
      R^{\bar{a}}{}_{\bar{b}} &= R^a{}_b.
  \end{align}
 \end{subequations}
Correspondingly after a Weyl transformation the spacetime changes as follows.  
 \begin{align*}
  \mathcal{W} : \quad 
    \begin{array}{lcl}
     \text{Minkowski}  & \longrightarrow  & \text{Symmetric teleparallel} \\
     \text{Riemann}  & \longrightarrow  & \text{Riemann-Weyl}\\
      \text{Weitzenböck teleparallel}  & \longrightarrow  & \text{General teleparallel}\\
              \text{Riemann-Cartan}  & \longrightarrow  & \text{Most general}\\
         \text{Symmetric teleparallel}  & \longrightarrow  & \text{Symmetric teleparallel}\\
         \text{Riemann-Weyl}  & \longrightarrow  & \text{Riemann-Weyl}\\
     \text{General teleparallel}  & \longrightarrow  & \text{General teleparallel}\\
      \text{Most general}  & \longrightarrow  & \text{Most general}
    \end{array}
\end{align*}
Then we realize that under a $\mathcal{W}$-transformation the geometry is not invariant in the first half, it is invariant in the second half. Thus if nature respects Lorentz symmetry and Weyl symmetry, the simplest geometry in which one can develop consistently alternative gravity models is the symmetric teleparallel spacetime.

In his original work \cite{weyl1918},\cite{weyl1919} Hermann Weyl introduced a new field which we name the Weyl potential 1-form, $B$. Thus one can define a new covariant exterior derivative of a geometrical object, $\mathcal{O}^\mu$, in the coordinate co-frame
 \begin{equation}
     \widehat{D} \mathcal{O}^\mu = d\mathcal{O}^\mu + \widehat{\omega}^\mu{}_\nu \wedge \mathcal{O}^\nu
 \end{equation}
where the Weyl full connection 1-form $\widehat{\omega}^\mu{}_\nu$ is written as
 \begin{equation}
     \widehat{\omega}^\mu{}_\nu = \omega^\mu{}_\nu + w \delta^\mu_\nu  B .
 \end{equation}
Here $w$ is the conformal weight of $\mathcal{O}^\mu$, i.e. $\mathcal{O}^{\bar{\mu}}=e^{w\psi(x)} \mathcal{O}^\mu$ and also $\mathcal{O}^{\mu'}= \Lambda^{\mu'}{}_{\nu} \mathcal{O}^\nu$. In order $ \widehat{D} \mathcal{O}^\mu$ to transform in the same way as $\mathcal{O}^\mu$, that is $\widehat{D} \mathcal{O}^{\mu'}= \Lambda^{\mu'}{}_{\nu}\widehat{D} \mathcal{O}^\nu $ and $\widehat{D} \mathcal{O}^{\bar{\mu}}= e^{w\psi(x)} \widehat{D} \mathcal{O}^\mu$ under both transformations of a local GCT and a Weyl transformation, we prescribe transformation rules for the Weyl 1-form as
 \begin{equation}
     B'=B \quad \text{and} \quad \bar{B}= B-d\psi(x) .
 \end{equation}
In his theory Weyl required that the covariant derivative satisfies the semimetricity condition, $\nabla_\sigma g_{\mu\nu}=-2 B_\sigma g_{\mu\nu}$. In our formalism and notation the Weyl equation corresponds to 
 \begin{equation}
     Q_{\mu\nu}= B g_{\mu\nu}.
 \end{equation}
This is invariant under both a GCT and a Weyl transformation. Weyl thought of $B$ as the Maxwell's electromagnetic potential 1-form and attempted to argue a unified theory of gravity and electromagnetism. Just after the publication it was seen that the Weyl potential 1-form does not couple the electric current, but to the dilaton current of matter. Indeed, there is no distinction between interactions of $B$ with particles and anti-particles and this is the oppose to observational data on electromagnetism. Meanwhile the first counter-argument to Weyl's theory was developed by Einstein. Einstein argued that Weyl's theory predicts a second clock effect (SCE) meaning that the tick-tack rate of a clock is dependent of its worldline (or history) which has been never observed. This theoretical effect is in addition to the well known first clock effect that is predicted both in special and general relativity and has been repeatedly observed in experiments. Basically a SCE originates from the geometric result that norm of a vector changes during parallel transport along its worldline (trajectory). Therefore, we search parallel transport of the tangent vector, $u^\mu=dx^\mu/d\lambda$, of a curve $\mathcal{C}$ defined by $x^\mu(\lambda)$ where $\lambda$ is an affine parameter below. With notation $u:=\left|u^\mu \right|$, the square of norm (or length) of $u^\mu(\lambda)$ is written via metric, $u^2 = g_{\mu\nu}u^\mu u^\nu$. Then we differentiate this with respect to the full connection, $\omega^\mu{}_\nu$,
 \begin{align}
     D u^2 = D(g_{\mu\nu}u^\mu u^\nu) \quad \Rightarrow \quad 
      udu = -Q_{\mu\nu} u^\mu u^\nu + g_{\mu\nu} u^\mu Du^\nu . \label{eq:lenthchange1}
 \end{align}
Now the critical question is ``what is the rule of parallel transport of a tangent vector to a curve along the curve?''. In the standard textbooks it is written as
 \begin{equation}
    D u^\mu=0 \quad \Rightarrow \quad 
     \frac{ D u^\mu}{d\lambda}=0  \quad \Rightarrow \quad 
      \left(\frac{ D u^\mu}{\partial x^\nu}\right) \left( \frac{dx^\nu}{d\lambda}\right)=0  \quad \Rightarrow \quad 
     u^\nu  D_\nu u^\mu=0 \label{eq:autoparallel1}
 \end{equation}
which turns out to be in components
 \begin{equation}
     \frac{d^2x^\mu}{d\lambda^2} + \omega^\mu{}_{\nu , \sigma} \frac{dx^\nu}{d\lambda} \frac{dx^\sigma}{d\lambda}  =0 \label{eq:autoparallel2}
 \end{equation}
where $\omega^\mu{}_{\nu , \sigma} := \iota_\sigma \omega^\mu{}_{\nu}$ or $\omega^\mu{}_{\nu} = \omega^\mu{}_{\nu , \sigma} dx^\sigma$. It is worthy to remind that the full connection $\omega^\mu{}_{\nu , \sigma}$ contains the Levi-Civita, non-metricity and torsion pieces through the equation (\ref{connec-decomp-coor}). The rule of parallel transport given by (\ref{eq:autoparallel1}) or (\ref{eq:autoparallel2}) is also called as the equation of autoparallel curve. Only in Riemannian spacetimes the equation of autoparallel curve is the same as the geodesic equation obtained from the extreme of the integral $\int_{\lambda_i}^{\lambda_f} \sqrt{-g_{\mu\nu}(x) (dx^\mu/d\lambda) (dx^\nu/d\lambda)} d\lambda$ which defines the length of a timelike curve between the fixed end points $x^\mu(\lambda_i)$ and $x^\mu(\lambda_f)$. Consequently, after this definition of parallel transport (\ref{eq:autoparallel1}), the equation (\ref{eq:lenthchange1}) becomes
 \begin{equation*}
     udu= - Q_{\mu\nu , \sigma} u^\mu u^\nu dx^\sigma .
 \end{equation*}
Here we define the unit vector of $u^\mu$ as $t^\mu= u^\mu / u$, then rewrite the equation and integrate it
 \begin{equation}
     \frac{du}{u} = - Q_{\mu\nu , \sigma} t^\mu t^\nu dx^\sigma \quad \Rightarrow \quad u = u_i e^{-\int_{\lambda_i}^{\lambda_f} Q_{\mu\nu , \sigma} t^\mu t^\nu dx^\sigma}
 \end{equation}
where $u_i$ is an integration constant corresponding to the initial length of the parallel transported vector and $u$ is its final length. This is the source of debate about SCE which we discuss below. In conclusion, because of existence of the non-metricity, in general, the length of a tangent vector changes during the parallel transport along a closed autoparallel curve. In the original Weyl theory, $Q_{\mu\nu}=Bg_{\mu\nu}$, this result yields
 \begin{equation}
     u = u_i e^{\oint_{\mathcal{C}} B_\mu dx^\mu}= u_i e^{\int_{\partial S} B} = u_i e^{\int_{S} dB}   
 \end{equation}
where we use firstly $g_{\mu\nu}t^\mu t^\nu=-1$ since the unit vector is timelike, and then the Stokes theorem. Here the closed curve $\partial S$ is the boundary of $S$ which is the region bounded by two different trajectories of a clock. In the literature $dB$ is called as the line curvature (or field strength) and denoted by $F=dB$. If $F$ does not vanish in the region $S$, even if the initial and final points of the clock are the same, then the tick-tack rates of the clock are to be different when it follows the different paths. This is known as SCE. If Weyl potential 1-form is an exact form, that is $B=d\varphi$ for any scalar function $\varphi(x)$, then automatically $F$ vanishes because of the Poincare lemma, $F=dB=d^2\varphi=0$, and SCE disappears completely. The case of $B=d\varphi$ is named as the Weyl integrable spacetime.

\section{Autoparallel curve in the coincident gauge of symmetric teleparallel geometry}

The geometry defined by the configuration $Q_{\mu \nu} \neq 0$, $T^\mu=0$, $R^\mu{}_\nu=0$ is the symmetric teleparallel geometry. The modified theories of gravity written in this geometry is called as the symmetric teleparallel gravity (STPG). This configuration gives three constraints  
 \begin{subequations} \label{eq:stpg1}
  \begin{align}
    -d g_{\mu \nu} + \omega_{\mu \nu}+\omega_{\nu \mu} &\neq 0 , \\
      {\omega^\mu}_\nu \wedge dx^\nu &= 0 , \\
      d {\omega^\mu}_\nu + {\omega^\mu}_\sigma \wedge {\omega^\sigma}_\nu &= 0 .
 \end{align}
 \end{subequations}
One can not solve ${\omega^\mu}_\nu$ analytically through these equations since the last one is not linear algebraic for it. But it could have been done in the Riemannian geometry, $Q_{\mu \nu} = 0$, $T^\mu=0$, $R^\mu{}_\nu \neq 0$. Nevertheless, one special solution to the equations (\ref{eq:stpg1})  for ${\omega^\mu}_\nu$ is to consider the case below
 \begin{equation}
   {\omega^\mu}_\nu = 0 \quad \Rightarrow \quad Q_{\mu \nu} = - \frac{1}{2} d g_{\mu \nu} \neq 0, \quad T^\mu =0 , \quad  {R^\mu}_\nu =0 .
 \end{equation}
This is called the coincident gauge\footnote{When we firstly realized this trick, we called it as {\it a gauge fixing} in our paper \cite{adak2013ijmpa}, later we read the nomenclature {\it coincident gauge} in the literature.}. Of course, one can pass to a different gauge (or coordinate system) via a local GCT (\ref{eq:gl-trans-connec-coord}) such that $ {\omega^{\mu'}}_{\nu'} \neq 0 $, but still $ Q_{\mu' \nu'} \neq 0 $, $T^{\mu'} =0$ and ${R^{\mu'}}_{\nu'} =0$. In the case of coincident gauge the standard rule of parallel transport of a tangent vector (\ref{eq:lenthchange1}) yields 
   \begin{equation}
       du^\mu =0 \quad \Rightarrow \quad \frac{d^2x^\mu}{d\lambda^2}=0 .
   \end{equation}
This is too far from representing the orbit of a spinless test particle in STPG. One attempt to overcome this problem is performed by two of us in the reference \cite{adak2022} in which a novel prescription for parallel transport of a vector was defined. However its Weyl invariance and SCE were not discussed. We realize that our original prescription (the equation (16) in \cite{adak2022}) and the standard prescription (the equation (\ref{eq:autoparallel1}) here) for parallel transport of a tangent vector are invariant under a local GCT, but not under a Weyl transformation. Consequently we give a new rule for parallel transport of a tangent vector, $u^\mu$, to a curve $x^\mu(\lambda)$
  \begin{equation} \label{eq:autopar3}
      Du^\mu=Q^\mu{}_\nu u^\nu
  \end{equation}
which is invariant under both a local GCT and a Weyl transformation. Also, this is free from SCE in general, not only for an integrable Weyl spacetime since the length of a vector needs not to change during parallel transport in spacetimes with non-metricity, $D\left|u^\mu\right|=0$. This new parallel transport equation corresponds to the set of $a=1$, $b=c=0$ in the equation (16) of \cite{adak2022}. In the Ref.\cite{hobson2020} the authors keep the length of a vector unchanged on completing a loop, and so remove the original basis suggesting the existence of an SCE by postulating a new metric compatibility condition and a rule for parallel transport of a vector
 \begin{equation}
     \widehat{D}g_{\mu\nu}=0 \quad \text{and} \quad \widehat{D} u^\mu=0 .
 \end{equation}
They stay invariant under both a local GCT and a Weyl transformation. Their equations may be expanded as
  \begin{subequations}
     \begin{align}
     \widehat{D}g_{\mu\nu} &= D g_{\mu\nu} + 2g_{\mu\nu} B=0 \quad \Rightarrow \quad Q_{\mu\nu}= g_{\mu\nu} B , \label{eq:hubson1} \\
     \widehat{D} u^\mu &= Du^\mu - B u^\mu=0 \qquad  \Rightarrow \quad Du^\mu= B u^\mu.
      \end{align}
 \end{subequations}
Thus their solution in order to terminate a SCE is valid only for a restricted components of the non-metricity: from the semicompatibility $B=Q/4 = g^{\mu\nu}Q_{\mu\nu}/4$, then the parallel transport $Du^\mu = Q u^\mu /4$. On the other hand, our suggestion (\ref{eq:autopar3}) is valid for all components of it. For example, a special choice of $Q_{\mu\nu}=g_{\mu\nu} Q/4$ in (\ref{eq:autopar3}) coincides with the result of \cite{hobson2020}. Similarly the author of \cite{israelquiros2022} bases his discussion on the assumption $B=\mathcal{Q}$ yielding
 \begin{equation}
     Q_{\mu\nu} = g_{\mu\nu} \mathcal{Q} = g_{\mu\nu} [a Q + (1-4a) \iota_\sigma Q^\sigma{}_\alpha dx^\alpha]
 \end{equation}
where $a$ is left as a free parameter. Since the choice of $a=1/4$ corresponds to (\ref{eq:hubson1}), it seems more general than the work \cite{hobson2020}, but still more restricted than our result (\ref{eq:autopar3}). We started our discussion with the coincident gauge of symmetric teleparallel spacetimes, but our result is general enough to cover any non-Riemannian geometry which consists of non-metricity and torsion.

One more remark is on the proper time, $\tau$. In GR it is defined as $ds^2=g_{\mu\nu} dx^\mu dx^\nu = -d\tau^2$ in natural units, $c=G=\hbar=1$ and the signature $(-,+,+,+)$. This definition is invariant under a local GCT, but not under a Weyl transformation due to (\ref{eq:weyl-trans-coor}). In the literature it is sometimes assumed that conformal weight of velocity four-vector to be zero, then $u^{\bar{\mu}}= u^\mu $, so $dx^{\bar{\mu}}/d\bar{\tau} = dx^\mu/d\bar{\tau}$ yielding a Weyl invariant proper time $d\bar{\tau}=d\tau$. But in this work $w(u^\mu)=-1$, i.e. $u^{\bar{\mu}}= e^{-\psi(x)}u^\mu$, and so $d\bar{\tau}=e^{\psi(x)}d\tau$. In order to have a Weyl invariant proper time we follow the work \cite{hobson2020},\cite{hobson2022} in which the arbitrary parameter $\lambda$ is not interpreted as the proper time of a particle moving along the worldline because one obtains a relation $d/d\bar{\lambda}=e^{-\psi(x)} d/d\lambda$ from the definition $u^\mu=dx^\mu / d\lambda$. In order to clarify the subject they first note that since Einstein’s objection to Weyl’s theory is based on the observation of sharp spectral lines, the presence of matter fields to represent atoms, observers and clocks is required. Then they consider Weyl gauge theories to include ordinary matter which is generally modeled by a Dirac field, see the equation (\ref{DirLag}). Finally they introduce a scalar compensator field\footnote{There is no relation between the compensator field $\phi$ and $\varphi$ in the Weyl potential 1-form $B=d\varphi$.} $\phi(x)$ with Weyl weight $w(\phi)=-1$, and conclude that an interval of proper time measured by the clock along its path is given by $d\tau = \phi d\lambda$. By definition this $d\tau$ is Weyl invariant. Accordingly, in scale-invariant gravity theories, they suggest that a compensator scalar field $\phi$ both enables dynamic generation of particle masses and is also key in generating a particle’s proper time. For more detailed discussion one can consult for  \cite{hobson2020},\cite{hobson2022}.  

We now wonder if our autoparallel transport recipe (\ref{eq:autopar3}) would represent a correct orbit equation of a test particle, e.g. the planet of Mercury. Therefore we write down the explicit form of it in the coincident gauge of symmetric teleparallel geometry as
  \begin{equation} \label{eq:autopar2}
      \frac{d^2x^\mu}{d\lambda^2} + \frac{1}{4} g^{\mu\beta} \left( \partial_\sigma g_{\beta \nu} + \partial_\nu g_{\beta \sigma}   \right) \frac{dx^\nu}{d\lambda} \frac{dx^\sigma}{d\lambda}=0 .
  \end{equation}
In terms of proper time this turns out to be
  \begin{equation} \label{eq:autopar4}
      \frac{d^2x^\mu}{d\tau^2} + \frac{1}{4} g^{\mu\beta} \left( \partial_\sigma g_{\beta \nu} + \partial_\nu g_{\beta \sigma}   \right) \frac{dx^\nu}{d\tau} \frac{dx^\sigma}{d\tau} + \frac{\partial_\nu \phi}{\phi}  \frac{dx^\nu}{d\tau} \frac{dx^\mu}{d\tau} =0.
  \end{equation}
Here we have calculated explicitly the components of this autoparallel curve equation for a spherically symmetric static metric, and experienced that our novel equation of an autoparallel curve can not represent orbit equation of a spinless test body, (e.g. even planet of Mercury).  So, we have to look at the geodesic equation whether it can be used for orbit equation of a spinless test body. In fact we ought to say meta-geodesic\footnote{Since nowadays metaverse is so popular, we prefer to use prefix ``meta'' instead of ``para'' or ``semi''!} equation, because Lorentz and Weyl invariant action contains the compensator field:  $I=\int \phi ds =\int_{i}^f \phi \sqrt{-g_{\mu\nu} dx^\mu dx^\nu}$. Then, $\delta I=0$ causes the meta-geodesic equation
 \begin{align}
     \frac{d^2x^\mu}{d\tau^2} + \frac{1}{2} g^{\mu\beta} \left( \partial_\sigma g_{\beta \nu} + \partial_\nu g_{\beta \sigma} - \partial_\beta g_{\nu \sigma}  \right) \frac{dx^\nu}{d\tau} \frac{dx^\sigma}{d\tau} + g^{\mu\nu} \partial_\nu \ln\phi   =0.
 \end{align}
Here since we make the length of tangent vector to be constant during parallel transport along the curve $x^\mu(\tau)$, we considered normalized 4-velocities, $\sqrt{-g_{\mu\nu} \frac{dx^\mu}{d\tau}\frac{dx^\nu}{d\tau}}=1$.
Now if we assume that the scalar field varies very slowly, then we can think of $\phi$ nearly constant in the scale of solar system, and so omit the last term: $\partial_\nu \phi \approx 0$. Consequently we obtain correct orbit equation of Mercury. But in galactic scale $\partial_\nu \phi$ may contribute a significant effect such that it may be responsible for the flatness of velocity curves at outer arms of spiral galaxies. So, this meta-geodesic equation may be the correct candidate for the worldline of a freely falling spinless test body.

\section{$SO(1,3)\otimes  \mathcal{W} \otimes U(1)$-invariant Lagrangian 4-form}

Finally we want to deal with the Weyl transformation of the quadratic parity preserving Lagrangian 4-form of symmetric teleparallel gravity given in the paper \cite{adak2013ijmpa} 
 \begin{align}
     L[Q] = c_1 & Q_{ab} \wedge *Q^{ab} + c_2 \left(Q_{ab} \wedge e^b\right) \wedge *\left(Q^{ac} \wedge e_c\right) + c_3 \iota^b Q_{ab} \wedge *\iota_c Q^{ac} \nonumber \\ 
     &+  c_4 Q \wedge *Q + c_5 \iota^b Q_{ab} \wedge * \iota^a Q  + \lambda_0 *1 + \lambda_a \wedge T^a + R^a{}_b \wedge \rho^b{}_a  \label{eq:STPGlagrangian1}
 \end{align}
where $c_1, \cdots , c_5$ are coupling constants, $\lambda_0$ is a constant which might be seen as cosmological constant, $\lambda_a$ is a Lagrange multiplier 2-form constraining torsion to zero, $\rho^b{}_a$ is another Lagrange multiplier 2-form terminating the full curvature. Since $T^{\bar{a}}=e^{\psi}T^a$ and $R^{\bar{a}}{}_{\bar{b}}=R^a{}_b$, we determine rule of Weyl transformations for the Lagrange multipliers as $\lambda_{\bar{a}}=e^{-\psi}\lambda_a$ and $\rho^{\bar{b}}{}_{\bar{a}}=\rho^b{}_a$. We prefer to work in the orthonormal co-frame because of user friendly and coordinate independent behavior of the Hodge dual star, $*1=e^0\wedge e^1 \wedge e^2 \wedge e^3$. Besides, one can consult for \cite{adak2018},\cite{adak2022ijgmmp} to see how the gauge spirit leads to that Lagrangian. It is invariant under the Lorentz transformation meaning that gauge group is $SO(1,3)$. Now we obtain the Weyl-transformed Lagrangian 4-form as
   \begin{align}
     \bar{L}[Q] =& e^{2\psi} \left[ c_1  Q_{ab} \wedge *Q^{ab} + c_2 \left(Q_{ab} \wedge e^b\right) \wedge *\left(Q^{ac} \wedge e_c\right) + c_3 \iota^b Q_{ab} \wedge *\iota_c Q^{ac} \right. \nonumber \\ 
     &\left.  + c_4 Q \wedge *Q + c_5 \iota^b Q_{ab} \wedge * \iota^a Q \right] + e^{4\psi}\Lambda *1 + \lambda_a \wedge T^a + R^a{}_b \wedge \rho^b{}_a \nonumber \\ 
    &-(2c_1 + 2c_2 + 8c_4 + c_5) Q\wedge *d\psi +  (2c_2 - 2c_3 - 4c_5) (\iota^aQ_{ab})e^b \wedge *d\psi \nonumber \\
    &+ ( 4c_1 + 3c_2 + c_3+ 16c_4 + 4c_5) d\psi \wedge *d\psi . \label{eq:STPGlagrangian2}
 \end{align}
As long as the relations among $c_i$s are valid
 \begin{subequations}\label{eq:conditions_on_cis}
  \begin{align}
      2c_1 + 2c_2 + 8c_4 + c_5 &=0 , \\
       2c_2 - 2c_3 - 4c_5 &=0 , \\
      4c_1 + 3c_2 + c_3+ 16c_4 + 4c_5 &=0 ,  
  \end{align}
 \end{subequations}
we can get rid of the residual terms. We notice here that two of these three algebraic relations are linearly independent and they are the same as the equation (44) of \cite{adak2013ijmpa}  and also GR-equivalent STPG given by the equation (31) of \cite{adak2013ijmpa} does not satisfy them. With help of compensator field $\bar{\phi}=e^{-\psi} \phi$ and the conditions (\ref{eq:conditions_on_cis}) the following symmetric teleparallel gravity Lagrangian 4-form is invariant under both Lorentz and Weyl transformations
   \begin{align}
     L[Q,\phi] =& \phi^2 \left[ c_1  Q_{ab} \wedge *Q^{ab} + c_2 \left(Q_{ab} \wedge e^b\right) \wedge *\left(Q^{ac} \wedge e_c\right) + c_3 \iota^b Q_{ab} \wedge *\iota_c Q^{ac} + c_4 Q \wedge *Q \right. \nonumber \\ 
     &\left.   + c_5 \iota^b Q_{ab} \wedge * \iota^a Q \right] + \mathcal{D}\phi \wedge * \mathcal{D}\phi + \lambda_0 \phi^4 *1 + \lambda_a \wedge T^a + R^a{}_b \wedge \rho^b{}_a  \label{eq:STPGlagrangian3}
 \end{align}
where $\lambda_0 \phi^4$ may be interpreted as mass term of the scalar field. Here again the gauge spirit makes us to introduce the kinetic term for $\phi$ in terms of the Weyl covariant exterior derivative of it 
 \begin{align}
     \mathcal{D}\phi = d\phi -\mathcal{Q}\phi \quad \text{where} \quad \mathcal{Q}= a Q +(1-4a)(\iota_aQ^{ab})e_b .
 \end{align}
Here $\mathcal{Q}$ is the general trace 1-form of non-metricity and $a$ is a free parameter. Now the gauge group is $SO(1,3)\otimes \mathcal{W}$.

As a last task we separately discuss the transformations of the Dirac-Maxwell Lagrangian 4-form
 \begin{equation}
 L[\Psi,A,\phi] = \text{Her} \left(i \overline{\Psi} *\gamma \wedge D\Psi \right) + im_0 \phi \,\overline{\Psi}\Psi \, *1 \, + dA \wedge *dA  \label{DirLag}
 \end{equation}
where $i=\sqrt{-1}$ is the imaginary unit, $m_0$ is a constant such that $m_0\phi$ represents the mass of Dirac field, $A$ is the Maxwell potential 1-form, $\gamma:=\gamma_a e^a$ is $\mathcal{C}\ell_{1,3}$-valued 1-form and $\overline{\Psi}$ is the Dirac adjoint of $\Psi$ spinor defined by $\overline{\Psi}:= \Psi^\dag \gamma_0$. Dirac matrices which are also known as the generators of the Clifford algebra $\mathcal{C}\ell_{1,3}$ must satisfy the relation $\gamma_a \gamma_b + \gamma_b \gamma_a = 2 \eta_{ab}I$ where $I$ is $4\times 4$ unit matrix that is not usually written explicitly. 4-component spinor fields $\Psi$ can be seen as sections from the base manifold to the spinor bundle. Its $SO(1,3)\otimes\mathcal{W}\otimes U(1)$-covariant  exterior derivative is in the form
 \begin{align}\label{eq:cov-der-Psi}
     D\Psi = d\Psi + \Omega \Psi
 \end{align}
where $\Omega$ is the full connection 1-form of the spinor bundle 
 \begin{align}
     \Omega = \frac{1}{2}\omega^a{}_b \Xi^b{}_a - \frac{1}{2}(3 + \Xi^a{}_a) \mathcal{Q} -iA \, .
 \end{align}
Here $\Xi^b{}_a$ contains some combinations of the Dirac matrices \cite{koivisto2020}. Since we already know transformation rules for $\omega^a{}_b$ and $\mathcal{Q}$ under both $SO(1,3)$ and $\mathcal{W}$, now we cast those of $A$ under $U(1)$ as $A \to A+df$ where $f(x)$ is a new scalar function and under  $\mathcal{W}$ as $A \to A$ meaning $A_{\bar{a}}=e^{-\psi}A_a$, and so $A_{\bar{\mu}}=A_\mu$. Then, as $\Psi \to e^{-3\psi/2} e^{if} S \Psi$, the equation (\ref{eq:cov-der-Psi}) transforms covariantly $D\Psi \to e^{-3\psi/2} e^{if} S D\Psi$, when $\Omega$ transforms according with the rule
 \begin{align}
     \Omega \to S \Omega S^{-1} + S dS^{-1} + \frac{3}{2} d\psi -i df .
 \end{align}
where $S$ is an element of $Spin(1,3)$ group that is the double cover of $SO(1,3)$. As a result $L[Q,\phi,A]= L[Q,\phi]+ dA \wedge *dA$ can be the true symmetric teleparallel theory combining gravitational and the electromagnetic interactions respecting the $SO(1,3)\otimes \mathcal{W} \otimes U(1)$-gauge symmetry. To obtain variational field equations and to search for their solutions will be discussed in another study in future.

\section{Conclusions}

In this work we aimed to concern the issues of second clock effect and proper time for toy models of gravity formulated in symmetric teleparallel (Minkowski-Weyl) geometry. Since we use extensively the exterior algebra, firstly we reviewed some basic concepts such as a coordinate system set on a manifold, coordinate and orthonormal frame bundles constructed over manifold, their connections (or covariant derivatives) and the passages among coordinate and orthonormal quantities via vielbein. Secondly we summarized a local general coordinate transformation causing the affine group on manifold, $Gl(4,\mathbb{R})$ group on coordinate frame bundle and $SO(1,3)$ group on orthonormal frame bundle, and then a Weyl transformation in terms of coordinate and orthonormal frames. Our first conclusion was that if nature obeys Weyl symmetry as well as Lorentz symmetry, then the unchanged simplest spacetime geometry before and after a Weyl transformation is the symmetric teleparallel geometry; $Q_{ab}\neq 0$, $T^a=0$, $R^a{}_b=0$. After discussing the change of length of a tangent vector, $u^\mu:=dx^\mu/d\lambda$, during parallel transport along a curve, $x^\mu(\lambda)$ where $\lambda$ is any parameter, we formulated a second clock effect depending on the existence of non-metricity. We saw that it can be avoided for a Weyl integrable spacetime as stated repeatedly in the literature. After that, we gave very briefly the coincident gauge of a symmetric teleparallel geometry by $\omega^\mu{}_\nu=0$ and wrote the standard equation of an autoparallel curve which could represent the orbit equation of a spinless test body. Thus, as our second conclusion we prescribed a novel parallel transport rule by (\ref{eq:autopar3}) which is invariant under a local GCT as well as a Weyl transformation. Accordingly, a second clock effect needs not to appear generally not only in a symmetric teleparallel geometry but also in a non-Riemannian spacetime with non-metricity and torsion. Meanwhile, we handled the notion of proper time and defined it in a $Gl(4,\mathbb{R})$ and Weyl invariant way with the help of a new compensator scalar field $\phi(x)$ by following \cite{hobson2020},\cite{hobson2022}. In the third conclusion we argued that the remedied equation of autoparallel curve can not still represent the worldline of a test body, but the modified equation of geodesic can be. In our final conclusion, we wrote down the Lagrangian 4-form of a symmetric teleparallel theory of gravity in terms of parity preserving quadratic non-metricity and the scalar field $\phi$, and also of Dirac-Maxwell all which are invariant under a $SO(1,3)\otimes \mathcal{W}\otimes U(1)$-transformation. That may be the true unified theory of gravitational and electromagnetic interactions. We left an open problem of obtaining variational field equations and of searching solutions to them for our future projects.

\section*{Acknowledgements}

%We thank to the anonymous referees for their guiding questions and criticisms.
In this study, C.P. and M.A. were supported via the project number 2022FEBE032 by the Scientific Research Coordination Unit of Pamukkale University.


\begin{thebibliography}{99}

\bibitem{weyl1918} H. Weyl, Gravitation und Elektrizität. Sitzungsber. Preuss. Akad. Berlin., pp. 465–480 (1918). Also as a chapter in the book Das Relativitätsprinzip, English translation at http://www.tgeorgiev.net/Gravitation\_and\_Electricity.pdf

\bibitem{weyl1919} H. Weyl, A new extension of the theory of relativity, {\it Annalen der Physik} {\bf 59} (1919) 101, doi: 10.1002/andp.19193641002

\bibitem{lobo2018} I. P. Lobo and C. Romero, Experimental constraints on the second clock effect, {\it Phys. Lett. B} {\bf 783} (2018) 306, doi: 10.1016/j.physletb.2018.07.019, arxiv: 1807.07188

\bibitem{benn1982} I. M. Benn, T. Dereli and R. W. Tucker, A critical analysis of some fundamental differences in gauge approaches to gravitation {\it J. Phys. A} {\bf 15} (1982) 849-866, doi: 10.1088/0305-4470/15/3/023

\bibitem{tucker1995} R. W. Tucker and C. Wang, Black holes with Weyl charge and non-Riemannian waves, {\it Class. Quantum Grav.} {\bf 12} (1995) 2587-2606, doi: 10.1088/0264-9381/12/10/016, arxiv: gr-qc/9509011

\bibitem{hehl1995}  F. W. Hehl {\it et al.}, Metric-affine gauge theory of gravity: field equations, Noether identities, world spinors, and breaking of dilation invariance, {\it Phys. Rep.} {\bf 258} (1995) 1, doi: 10.1016/0370-1573(94)00111-F, arxiv: gr-qc/9402012

\bibitem{adak2004_1} M. Adak and T. Dereli, Possible effects of spacetime nonmetricity on neutrino oscillations, {\it Phys. Rev. D} {\bf 69} (2004) 123002, doi: 10.1103/PhysRevD.69.123002, arxiv: gr-qc/0303080

\bibitem{adak2006}   M. Adak, M. Kalay and O. Sert, Lagrange formulation of the symmetric teleparallel gravity, {\it Int. J. Mod. Phys. D} {\bf 15} (2006) 619-634, doi: 10.1142/S0218271806008474, arxiv: gr-qc/0505025

\bibitem{Adak2006_1} M. Adak, The Symmetric Teleparallel Gravity, {\it Turk. J. Phys.} {\bf 30} (2006) 379-390, doi: 10.3906/zoo-1205-26, arxiv: gr-qc/0611077

\bibitem{adak2008} M. Adak and T. Dereli, The quadratic symmetric teleparallel gravity in two dimensions, {\it EPL} {\bf 82} (2008) 30008, doi: 10.1209/0295-5075/82/30008, arxiv: hep-th/0607058

\bibitem{demir2012} C. N. Karahan, A. Altas and D. A. Demir, Scalars, vectors and tensors from metric-affine gravity, {\it Gen. Relativ. Gravit.} {\bf 45} (2013) 319, doi: 10.1007/s10714-012-1473-x, arxiv: 1110.5168 

\bibitem{adak2013ijmpa} M. Adak, O. Sert, M. Kalay and M. Sari, Symmetric Teleparallel Gravity: Some exact solutions and spinor couplings, {\it Int. J. Mod. Phys. A} {\bf 28} (2013) 1350167, doi: 10.1142/S0217751X13501674, arxiv: 0810.2388

\bibitem{delhom2017} A. D. Latorre, G. J. Olmo and M. Ronco, Observable traces of non-metricity: New constraints on metric-affine gravity, {\it Phys. Lett. B} {\bf 780} (2018) 294, doi: 10.1016/j.physletb.2018.03.002, arxiv: 1709.04249

\bibitem{Tomi2018} T. Koivisto, An integrable geometric foundation of gravity, {\it Int. J. Geom. Methods Mod. Phys.} {\bf 15} (2018) 1840006, doi: 10.1142/S0219887818400066, arxiv: 1802.00650

\bibitem{adak2018} M. Adak, Gauge approach to the symmetric teleparallel gravity, {\it Int. J. Geomet. Meth. Mod. Phys.} {\bf 15} (2018) 1850198, doi: 10.1142/S0219887818501980, arxiv: 1809.01385

\bibitem{adak2022ijgmmp} C. Pala, E. Kok, O. Sert, M. Adak, A modified gravity model coupled to a Dirac field in 2D spacetimes with quadratic nonmetricity and curvature,  {\it Int. J. Geomet. Meth. Mod. Phys.}
{\bf 19} (2022) 2250045, doi: 10.1142/S0219887822500451, arxiv: 2011.10982 

\bibitem{jzyang2021} J-Z. Yang et al., Geodesic deviation, Raychaudhuri equation, Newtonian limit, and tidal forces in Weyl-type f(Q,T)f(Q,T) gravity, {\it Eur. Phys. J. C} {\bf 81} (2021) 111, doi: 10.1140/epjc/s10052-021-08910-6, arxiv: 2101.09956

\bibitem{albuquerque2022} I. Albuquerque and N. Frusciante, A designer approach to f(Q) gravity and cosmological implications, {\it Phys. Dark Univ.} {\bf 35} (2022) 100980, doi: 10.1016/j.dark.2022.100980, arxiv: 2202.04637

\bibitem{jimenez2022} J. B. Jimenez and T. S. Koivisto, Lost in translation: the Abelian affine connection (in the coincident gauge), Contribution to: GeomGrav2021, {\it Int. J. Geomet. Meth. Mod. Phys.} {\bf 19} (2022) 2250108, doi: 10.1142/S0219887822501080, arxiv: 2202.01701

\bibitem{hohmann2022} K. Flathmann and M. Hohmann, Parametrized post-Newtonian limit of generalized scalar-nonmetricity theories of gravity, {\it Phys. Rev. D} {\bf 105} (2022) 044002, doi: 10.1103/PhysRevD.105.044002, arxiv: 2111.02806

\bibitem{wang2022} W. Wang, H. Chen and T. Katsuragawa, Static and spherically symmetric solutions in f(Q) gravity, {\it Phys. Rev. D} {\bf 105} (2022) 024060, doi: 10.1103/PhysRevD.105.024060, arxiv: 2110.13565

\bibitem{obukhov2022} Y. N. Obukhov,  Conservation laws in gauge gravity theory, {\it Int. J. Geom. Methods Mod. Phys.}, {\bf 19} (2022) 2240002, doi: 10.1142/S0219887822400023, arxiv: 2202.06112

\bibitem{adak2022} C. Pala and M. Adak, A novel approach to autoparallels for the theories of symmetric teleparallel gravity, {\it J. Phys.: Conf. Ser.} {\bf 2191} (2022) 012017, doi: 10.1088/1742-6596/2191/1/012017, arxiv: 1102.1878

\bibitem{ehlers1972} J. Ehlers, F. Pirani, and A. Schild (1972). In General Relativity, L. O'Raifeartaigh, ed. (Oxford U. P., New York). Republication: {\it Gen Relat. Grav.} (2012) {\bf 44} 1587, doi: 10.1007/s10714-012-1353-4

\bibitem{perliek1987} V. Perlick, Characterization of Standard Clocks by Means of Light Rays and Freely Falling Particles, {\it Gen. Relat. Grav.} {\bf 19} (1987) 1059, doi: 10.1007/BF00759142

\bibitem{tucker1996} P. Teyssandier and R. W. Tucker, Gravity, gauges and clocks, {\it Class. Quantum Grav.} {\bf 13} (1996)  145, doi: 10.1088/0264-9381/13/1/013, arxiv: gr-qc/9510045

\bibitem{tucker1998} P. Teyssandier, R. W. Tucker and C. Wang, On an interpretation of non-Riemannian gravitation, {\it Acta Phys. Pol. B} {\bf 29} (1998) 987, https://www.actaphys.uj.edu.pl/R/29/4/987/pdf

\bibitem{avalos2018} R. Avalos, F. Dahia and C. Romero, A Note on the Problem of Proper Time in Weyl Space–Time, {\it Found. Phys.} {\bf 48} (2018) 253, doi: 10.1007/s10701-017-0134-z, arxiv: 1611.10198

\bibitem{koivisto2020} J. B. Jimenez, L. Heisenberg and T. Koivisto, The coupling of matter and spacetime geometry,  {\it Class. Quant. Grav.} {\bf 37} (2020) 195013, doi: 10.1088/1361-6382/aba31b, arxiv: 2004.04606

\bibitem{hobson2020} M. P. Hobson and A. N. Lasenby, Weyl gauge theories of gravity do not predict a second clock effect   {\it Phys. Rev. D} {\bf 102} (2020)  084040, doi: 10.1103/PhysRevD.102.084040, arxiv: 2009.06407

\bibitem{hobson2022} M. P. Hobson and A. N. Lasenby, Note on the absence of the second clock effect in Weyl gauge theories of gravity, {\it Phys. Rev. D} {\bf 105} (2022) L021501,  doi: 10.1103/PhysRevD.105.L021501, arxiv: 2112.09967

\bibitem{adelhom2020} A. Delhom {et al.}, Conformally invariant proper time with general non-metricity,  {\it Eur. Phys. J. C} {\bf 80} (2020) 415, doi: 10.1140/epjc/s10052-020-7974-y, arxiv: 2001.10633

\bibitem{quiros2021} I. Quiros, Nonmetricity theories and aspects of gauge symmetry, {\it Phys. Rev. D} {\bf 105} (2022) 104060, doi: 10.1103/PhysRevD.105.104060, arxiv: 2111.05490

\bibitem{israelquiros2022} I. Quiros, Gauge invariant approach to nonmetricity theories and the second clock effect, arxiv: 2201.03076

\bibitem{thirring1997} W. Thirring, {\bf Classical Mathematical Physics: Dynamical Systems and Field Theories} (Springer-Verlag, 3rd Edition, 1997)

\bibitem{frankel2012} T. Frankel, {\bf The Geometry of Physics} (Cambridge University Press, 3rd Edition, 2012)


\end{thebibliography}
\end{document}